# Characterization of Terahertz Spectral Bands for Next Generation of Wireless Communications

Eeswar Kumar Yalavarthi, *Graduate Student Member, IEEE,* Wei Cui, Aswin Vishnuradhan, Nicolas Couture, Markus Betz, Angela Gamouras, Jean-Michel Ménard

*Abstract*—The ever-increasing demand for high-speed data transmission continues to motivate research and development efforts towards the sixth generation (6G) of wireless communication technologies and beyond. The use of terahertz (THz) carrier frequencies is considered to achieve faster data transmission rates, with the potential to reach terabits per second. However, there is a necessity to study the impact of environmental factors affecting the signal transmission at these frequencies. One key challenge of THz signal propagation through air is water vapor absorption, which can severely limit the broadcast distance of THz wireless communications. In this work, we investigate the propagation distance of several THz channels under 35 % relative humidity atmospheric conditions using a table-top THz spectroscopy apparatus equipped with a sensitive THz detection system based on nonlinear parametric upconversion. Our results show seven transmission bands between 1 and 3 THz with negligible signal absorption at different propagation lengths with potential application in short-, mid- and long-range wireless communications.

## I. Introduction

THE last few decades have seen a vast increase in wireless data traffic as users across the globe consume data for communications and the internet of things (IoT) connected devices. Wireless data rates have experienced unprecedented growth, doubling every eighteen months over the past three decades [1]. Since the 1980's, wireless communication systems have evolved from speeds of approximately 2 kilobits per second to speeds of about 10 gigabits per second in the present fifth generation (5G) networks [2]. In 5G systems, millimeter-waves with frequencies below 100 gigahertz (GHz) restrict the contiguous bandwidth to 10 GHz, which is an important factor that determines data transfer rates.

To further improve data transfer rates, the sixth generation of wireless networks (6G) is expected to rely on frequencies above 100 GHz and up to 3 terahertz (THz) [3]. This largely untapped spectral window could support data rates of up to 1 terabit per second [2], [3], accommodating the growing number of connected devices, including autonomous systems requiring high data transfer rates. One major impediment to the implementation of THz wireless communications is the limited availability of efficient THz generation and detection systems. Although THz technologies are widely used in areas including astronomy, imaging, medicine and security [4], there remains a gap in sensitive, high-speed THz detection technologies. This limitation is exacerbated by the absorption of THz radiation propagating in normal atmospheric conditions due to water vapor, posing a major obstacle for long-distance THz wireless communication. However, low-absorption spectral windows can be identified to enable a range of applications relying on wireless signal transmission. Such spectral bands below 1 THz have been well explored [5], [6], and four possible data transmission bands between 1 – 2 THz have also been identified [7]. These studies were conducted using traditional THz time-domain spectroscopy (THz-TDS) systems which have high detection sensitivity and versatility in many applications in physics, chemistry, biology and engineering. However, THz-TDS systems are limited by their lengthy data acquisition times for single frequencies as they require numerous data points in the time domain to fully capture spectral information. These systems are also impractical for long-distance THz propagation distances due to the need for a stable temporal overlap between the THz pulse and a femtosecond gating pulse. Specifically, a 1 ºC variation in air temperature over a 50-meter distance can shift a 3 THz signal by a half cycle, making the technique ineffective.

In this work, we take advantage of a 10 ps near-infrared (NIR) gating pulse and a nonlinear upconversion process, sum frequency generation (SFG), inside a nonlinear optical crystal to detect THz radiation and to characterize the transmission spectrum. A NIR-sensitive single-photon avalanche diode (SPAD) operating at room-temperature is used to achieve THz detection sensitivity at the zeptojoule level [8]. We identify seven low-absorption spectral windows between 1 and 3 THz with a 100 GHz bandwidth. For each of these regions of

This work was supported by the High Throughput and Secure Networks Challenge Program at the National Research Council of Canada (HTSN 702, HTSN 254), and the Joint Centre for Extreme Photonics. *(Corresponding author: Jean-Michel Ménard)*

Eeswar K. Yalavarthi, Wei Cui, Aswin Vishnuradhan and Nicolas Couture are with the Department of Physics, University of Ottawa, Ottawa, ON, K1N 6N5, Canada (e-mail: eyala082@uottawa.ca, wcui065@uottawa.ca, avish009@uottawa.ca, ncout007@uottawa.ca).

Markus Betz is with the Department of Physics, TU Dortmund University, Dortmund, 44227, Germany and the Department of Physics, University of Ottawa, Ottawa, Ontario, K1N 6N5, Canada (e-mail: markus.betz@tu-dortmund.de).

Angela Gamouras and Jean-Michel Ménard are with the Department of Physics, University of Ottawa, Ottawa, Ontario, K1N 6N5, Canada and the National Research Council Canada, Ottawa, Ontario, K1A 0R6, Canada (e-mail: angela.gamouras@nrc-cnrc.gc.ca, jean-michel.menard@uottawa.ca).

interest (ROIs), we characterize the signal transmission in a 35% relative humidity atmosphere using four propagation distances between 50 and 125 cm. To assess the practical application of these ROIs in wireless communications, we extrapolate the signal strength over extended propagation distances.

## II. THz Measurement Apparatus

Figure 1 shows the experimental apparatus, which includes a Yb:KGW laser system (central wavelength: $\lambda_L = 1.035$ μm) delivering 180 fs pulses at a 50 kHz repetition rate. This laser beam is divided with a beam splitter (10% reflection/ 90% transmission) where the reflected beam is focused onto a 2 mm-thick gallium phosphide (GaP) crystal with a phase grating on its incident surface for efficient broadband THz generation [9], [10]. The transmitted beam is used as a gating pulse for THz detection. The general detection scheme is reported in our previous work [8]. In brief, the THz detection concept relies on a nonlinear optical interaction involving the THz pulse and the NIR gating pulse inside a GaP crystal, leading to upconverted NIR photons.

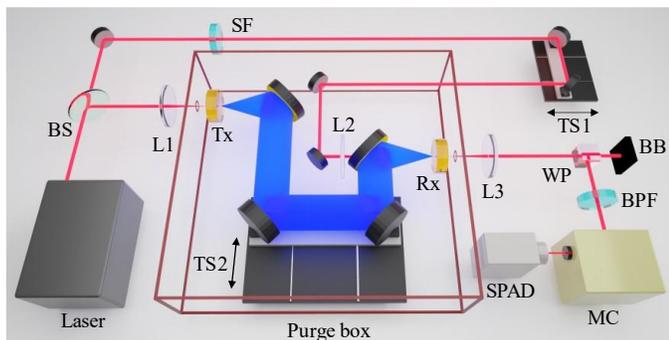

**Fig. 1.** Schematic diagram of the table-top apparatus. The THz propagation distance is varied using a translation stage (TS2) and enables the measurement of THz signals at different propagation lengths. Laser: $\lambda_L = 1035$ nm, BS: Beam splitter, TS1 & TS2: translation stages, L1, L2 and L3: lens, Tx (transmitter): 2mm-thick gallium phosphide crystal with grating [9], [10], Rx (receiver): 2mm-thick gallium phosphide crystal, WP: Wollaston prism, SF: spectral filters, BPF: Band-pass filters, MC: monochromator, SPAD: Single-photon avalanche diode, BB: Beam block, Purge box: Humidity controlled chamber which can be set to relative humidity of 35% and 0%.

A monochromator and a SPAD combination then monitor the upconverted signal, from which we can recover the THz pulse energy corresponding to a fixed spectral bandwidth. For this experiment, the laser beam is transmitted through spectral filters that decrease the bandwidth of the gating pulse to $\Delta\lambda = 0.15$ nm at full width at half maximum (FWHM) in the frequency domain, which leads to 10.4 ps pulse duration in the time domain. Considering these parameters and the resolution of the monochromator, our detection system achieves a resolution of 0.1 THz. A translation stage allows for the propagation distance of the THz signal to be varied from 52 cm to 124 cm. The experiment setup is enclosed in a purge box, allowing us to alternate between humidity levels of 35% and 0%.

## III. Results and Discussion

The THz spectrum, corresponding to the SFG signal, is recorded for propagation lengths of 52, 76, 103 and 124 cm in both dry air (0% relative humidity) and typical (35% relative humidity) environments at 21 ºC. The impact of water vapor on the THz spectrum is determined by taking the ratio of these two measurements for each propagation distance. The measured spectral transmission, along with high-resolution water vapor transmission spectra taken from the literature [11], is shown in Fig. 2.

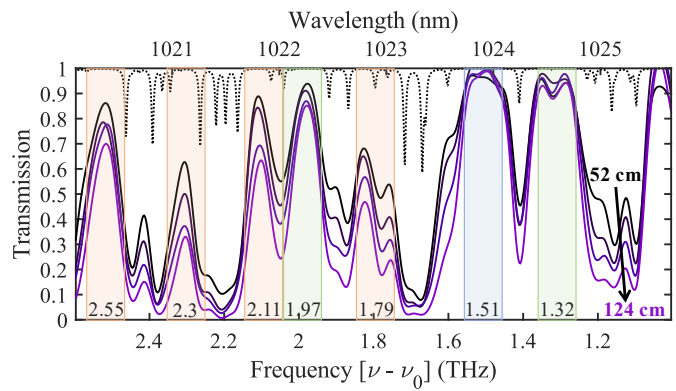

**Fig. 2.** Measured THz spectral transmission for four beam propagation distances (52, 76, 103 and 124 cm) in a 35% relative humidity environment. The line color gradient, from black to purple, follows the increase in THz propagation distances. The dotted line indicates the water vapor transmission obtained from HITRAN database [11], where relative absorption strength is set to arbitrary units for better comparison with experimental data. Different ROIs are relevant for applications in short-range (pale orange), mid-range (pale green) and long-range (pale blue) communications.

Seven spectral windows of relatively high transmission can be identified between the water vapor absorption lines centered at 1.32, 1.51, 1.79, 1.97, 2.11, 2.3 and 2.55 THz. Each ROI has a spectral bandwidth of 100 GHz corresponding to the spectral resolution of our experimental setup. We classify these ROIs into three color-coded categories, shown in Fig. 2, corresponding to different types of applications. Pale orange is used to identify the carrier frequencies practical for short-range communication ($\leq$ 20 m). The ROIs with potential for mid-range communication ($>$ 20 and $<$ 200 m) are in pale green and pale blue indicates the ROI at 1.51 THz, which can support relatively long-range ($>$ 200 m) communications.

The measured transmission data for each of the seven windows at four THz propagation distances are shown in Fig. 3. The error bars indicate the standard deviation of the total calculated error, where the measurement uncertainties are estimated based on fluctuations in the experiment set-up and local humidity variations. Within the measurement apparatus, the sources of error include laser power fluctuations and background noise detected by the SPAD. At each of the four distances, we take the mean of 300 data points, averaging over 50,000 pulses for each data point, measured in both humid and dry air conditions. Based on a precise knowledge of the transmission and its error at four propagation distances up to 124 cm, we calculate the weighted mean (black line in Fig. 3) from Beer's law. This allows us to extrapolate the transmission to longer propagation distances within a defined confidence interval. This confidence interval is determined by calculating the error in transmission at an arbitrary distance $z$ using the following approach. First, transmission coefficients $T_i$, measured at the four investigated propagation distances $z_i$, are used to obtain corresponding absorption coefficients $\alpha_i$ using the relation:

$$T_i = e^{-\alpha_i z_i} \quad (1)$$

The transmission corresponding to the top and bottom of the error bars of each measurements are then plugged in the same equation to obtain two absorption coefficients, $\alpha_{i,+}$ and $\alpha_{i,-}$, from which a corresponding transmission error $\Delta T_i(z)$ is obtained from Beer's law for an arbitrary distance $z$. The weighted mean transmission [12] is then calculated using:

$$\bar{T}_w = \left(\sum_{i=1}^{4} \frac{T_i}{(\Delta T_i)^2}\right) / \left(\sum_{i=1}^{4} \frac{1}{(\Delta T_i)^2}\right) \quad (2)$$

A similar approach is used to compute the weighted standard deviation (purple-shaded region), obtained by replacing $T_i$ by $\Delta T_i$ in Eq. (2). Finally, the standard error (red-shaded region) is calculated by dividing the weighted standard deviation with effective number of measurements [12]. The results of this analysis are shown in Fig. 3 (b).

We evaluate a maximum propagation distance based on the minimum signal that can be resolved in real time by our detection technique, indicated by the grey dashed line in Fig. 3(b). This limit is mainly set by the upconversion efficiency inside the nonlinear crystal and the intensity of the THz source, which delivers pulses containing ~0.2 fJ of energy within each ROI. We set the detection limit at a signal-to-noise ratio of 1, corresponding to a transmission of $3 \times 10^{-4}$ or a detected THz pulse energy of ~60 zJ. Based on this limit, the maximum propagation distance of the THz signal contained within each ROI is calculated and is shown in Table I. For the ROIs centered at 1.79, 2.3, 2.55 and 2.11 THz, we find propagation distances to be approximately 5, 10, 15 and 20 m, respectively. Those at 1.32 and 1.97 THz yield THz propagation distances of 45 and 50 m, respectively. The last ROI, centered at 1.51 THz, yields a propagation distance of 210 m.

The seven ROIs identified in this work have the potential to be used for wireless communications over different distance ranges. The first category of ROI (pale orange in Fig. 2 and Table I) can be used for short-range communications with a THz propagation distance of ≤ 20 m. Related applications include secure communications within a close range, such as high-speed wireless information transfer from a laptop to a mobile phone, or direct file transfers in a meeting room, or kiosk downloading [13]. The second category of ROI (pale green in Fig. 2 and Table I), with a THz propagation distance of approximately 50 m, could be used in wireless local area network (WLAN) systems for high-speed high-volume data transfer in public area such as airports, conference centers, trains [14] and planes. The third category of ROI (pale blue in Fig. 2 and Table I) has a THz propagation distance > 200 m, with prospective implementations in single hop signal transmission across water bodies such as rivers and lakes, and in 6G mobile services with mobile towers across streets and motorways [2], [15].

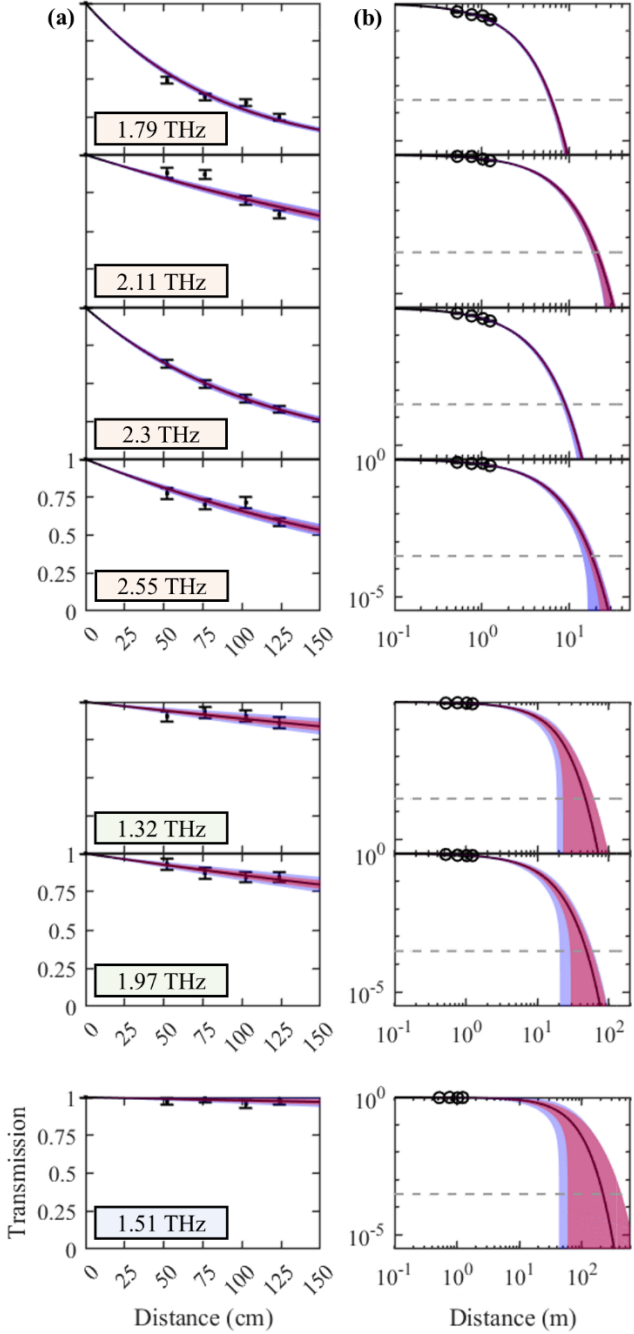

**Fig. 3.** (a) Measured transmission for seven ROIs at THz propagation lengths of 52, 76, 103 and 124 cm (black dots). The weighted mean (black line) is computed from the experimental data (black dots). The purple-shaded region indicates the weighted standard deviation and red-shaded region indicates the weighted standard error. (b) The THz transmission can be extrapolated to determine the maximum operational distance when the signal reaches the detection limit of our configuration (grey dashed line).

TABLE I
SUMMARY OF THZ RADIATION PROPAGATION AND APPLICATIONS

| Frequency (THz) | Distance limit (m) | Application |
|---|---|---|
| 1.79 | 5 | Short-range communications |
| 2.3 | 10 | Short-range communications |
| 2.55 | 15 | Short-range communications |
| 2.11 | 20 | Short-range communications |
| 1.32 | 45 | Mid-range communications |
| 1.97 | 50 | Mid-range communications |
| 1.51 | 210 | Long-range communications |

The propagation distances reported in this study are, in part, determined by the configuration of our experiment setup. Significantly longer THz operational distances can be achieved, for instance by increasing the emitted THz pulse energy and/or by further increasing the detector sensitivity. The latter could potentially be achieved using a structured nonlinear crystal [9], [16] to increase the THz to NIR conversion efficiency.

V. CONCLUSION

Using a tabletop THz spectroscopy apparatus, we investigate THz transmission between 1 and 3 THz under a 35% relative humidity environment. We identify seven 100 GHz-width frequency bands yielding significantly long transmission ranges for wireless signal propagation. These bands could be used for short-, mid- or long-range THz communication. Besides water vapor absorption, there are other technical challenges related to the development and realization of THz wireless communications and future 6G communications systems, many of those are not considered in this study. These include signal intensity reduction due to beam divergence, beam tracking requirements due to the signal directionality, multipath interferences in close space, and real-time fluctuations of environmental conditions. However, continuous progress in THz detection and emission technologies, including sensitive nonlinear detectors and high-power quantum cascade lasers, could lead to wireless THz systems with effective propagation distances exceeding those reported here.